\documentclass[letterpaper,twocolumn,10pt]{article}
\usepackage{usenix2019,epsfig}
\usepackage{cite}
\usepackage{graphicx}
\usepackage[hyphens]{url}
\usepackage{booktabs}
\usepackage{todonotes}
\usepackage{threeparttable}
\usepackage{comment}
\graphicspath{{./figures/}}
\usepackage{changes}
\usepackage{titlesec}
\usepackage{enumitem}

\titlespacing{\subsection}{0pt}{2ex}{1ex}
\titlespacing{\paragraph}{0pt}{1ex}{\wordsep}
\setlist{itemsep=0pt}

\begin{document}

\date{}

\title{\Large \bf Evaluating the Contextual Integrity of Privacy Regulation:\\Parents' IoT Toy Privacy Norms Versus COPPA}

\author{
{\rm Noah Apthorpe}\\
Princeton University
\and 
{\rm Sarah Varghese} \\
Princeton University
\and
{\rm Nick Feamster} \\
Princeton University
} 

\maketitle

\pagestyle{empty}

\subsection*{Abstract}
\noindent Increased concern about data privacy has prompted new and updated data protection regulations worldwide.
However, there has been no rigorous way to test whether the practices mandated by these regulations actually align with the privacy norms of affected populations. 
Here, we demonstrate that surveys based on the theory of contextual integrity provide a quantifiable and scalable method for measuring the conformity of specific regulatory provisions to privacy norms.
We apply this method to the U.S. Children's Online Privacy Protection Act (COPPA), surveying 195 parents and providing the first data that COPPA's mandates generally align with parents' privacy expectations for Internet-connected ``smart'' children's toys.
Nevertheless, variations in the acceptability of data collection across specific smart toys, information types, parent ages, and other conditions emphasize the importance of detailed contextual factors to privacy norms, which may not be adequately captured~by~COPPA.

\section{Introduction}

Data privacy protections in the United States are enforced through a combination of state and federal legislation and regulatory action.
In Europe, the General Data Protection Regulation (GDPR) is currently the best example of strong, centralized privacy legislation. The GDPR has inspired similar laws in other countries, such as the Brazilian General Data Privacy Law. According to the United Nations Conference on Trade and Development~\cite{UNCTD}, 57\% of countries have data protection and privacy legislation as of 2018. 

Although data privacy protections vary across countries in terms of details and implementation, many share a common provenance: public pressure to protect sensitive personal data from unauthorized use or release.
Surveys report that consumers worldwide were more concerned about online privacy in 2016 than 2014~\cite{IPSOS} and that over 60\% of U.S. survey respondents in 2018 are concerned about data privacy in general~\cite{marketwatch}. 
However, there has  been no rigorous, quantifiable, and scalable way to measure whether existing legal privacy protections actually match the privacy expectations of affected individuals.
Without such data, it is difficult to know which aspects of privacy regulation effectively align company behaviors with social and cultural privacy norms and which necessitate further revision.  

In this paper, we demonstrate that an existing survey technique \cite{noah} based on the formal privacy theory of contextual integrity (CI)~\cite{nissenbaum2004privacy} can be directly adapted to test the conformity of specific regulatory requirements to privacy norms, providing much-needed data to policymakers and the privacy research community.
The survey technique 
can be applied to any privacy regulation that defines guidelines for data collection and transfer practices. Importantly, the survey technique involves questions describing privacy scenarios that are concrete and understandable to respondents from all backgrounds. It also allows straightforward longitudinal and cross-sector measurements to track the effectiveness of regulatory updates over time.

We present a rigorous case study of this technique evaluating
the U.S. Children's Online Privacy Protection Act (COPPA), which provides a federal legal framework to protect the online privacy of children under the age of 13.  Specifically, we investigate whether parents' opinions about the acceptability of data collection practices by Internet-connected ``smart'' children's toys match COPPA mandates. Since the Federal Trade Commission (FTC) only updated its guidance on COPPA  to explicitly include ``connected toys or other Internet of Things devices'' in June 2017~\cite{futureical}, our results provide the first indication as to whether COPPA aligns with parents' privacy expectations.

This question is particularly relevant given the recent high-profile security breaches of smart toys,
ranging from the theft of personal information of over 6~million children from toy manufacturer VTech to vulnerabilities in Mattel's Hello Barbie \cite{finkle_bartz_patnaik_2015}. More recently, Germany banned children's smart watches and Genesis Toys' My Friend Cayla doll, citing security risks and ``spying concerns'' \cite{nienaber;_2017,techcrunch_2017}.

We survey a panel of 195 U.S.~parents of children from ages 3 to 13, the largest sample size for a study of parent opinions of smart toy data collection in the literature to date. 
We find that parents generally view information collection predicated on requirements specified by COPPA (e.g., ``if the information is used to protect a child's safety'') as acceptable, while viewing equivalent information collection without COPPA-specified conditions as unacceptable. This indicates that the existing conditions COPPA places on information collection by smart toys are generally in line with parents' privacy norms, although there may be additional data collection requirements which could be added to regulation that were not tested in our study. 

Additionally, we find that COPPA requirements for notification and consent 
result in more acceptable data collection practices than requirements related to confidentiality and security. This corroborates previous work indicating the primary importance of consent to user privacy norms~\cite{noah}. We also find variations in the acceptability of COPPA-permitted data collection practices across specific smart toys, types of information, certain information use cases, parent ages, parent familiarity with COPPA, and whether parents own smart devices. These variations emphasize the importance of detailed contextual factors to parents' privacy norms and motivate additional studies of populations with privacy norms that may be poorly represented by COPPA.

We conclude by noting that COPPA's information collection criteria are broad enough to allow smart toy implementations that compromise children's privacy while still adhering to the letter of the law. Continuing reports of smart toys violating COPPA~\cite{chu2018security} also suggest that many non-compliant toys remain available for purchase. 
Further improvements to both data privacy regulation and enforcement are still needed to keep pace with corporate practices, technological advancements, and privacy norms.

In summary, this paper makes the following contributions:
\begin{itemize}
    \item Demonstrates that an existing
    survey method \cite{noah} based on contextual integrity \cite{nissenbaum2004privacy} can be applied to test
    whether privacy regulations effectively match the norms of affected populations. 
    \item Provides the first quantitative evidence that COPPA's restrictions on smart toy data collection generally align with parents' privacy expectations.
    \item Serves as a template for future work using contextual integrity surveys to analyze current or proposed privacy regulation for policy or systems design insights.
\end{itemize}
\vspace{8pt}

\section{Background \& Related Work}

In this section, we place our work in the context of related research on contextual integrity, COPPA, and smart toys.

\subsection{Contextual Integrity}
\label{sec:CI}

The theory of contextual integrity (CI) provides a well-established framework for studying privacy norms and expectations \cite{nissenbaum2004privacy}. 
Contextual integrity defines privacy as the appropriateness of information flows based on social or cultural norms in specific contexts. 
CI describes information flows using five parameters: (1) the subject of the information being transferred, (2) the sender of this information, (3) the attribute or type of information, (4) the recipient of the information, and (5) the transmission principle or condition imposed on the transfer of information from the sender to the recipient. 
For example, one might be comfortable with a search engine (\textit{recipient)} collecting their (\textit{subject \& sender)} Internet browsing history (\textit{attribute)} in order to improve search results (\textit{transmission principle}), but not in order to improve advertisement targeting, which is a different transmission principle that places the information in a different context governed by different norms. 
Privacy norms can therefore be inferred from the reported appropriateness and acceptability of information flows with varying combinations of these five parameters.

Previous research has used CI to discover and analyze privacy norms in various contexts. 
In 2012, Winter used CI to design an interview study investigating Internet of things (IoT) device practices that could be viewed as privacy violations \cite{winter2012privacy}. 

In 2016, Martin and Nissenbaum conducted a survey with vignette questions based on CI to understand discrepancies between people's stated privacy values and their actions in online spaces \cite{martin2016measuring}. Rather than straightforward contradictions, they find that these discrepancies are due to nuanced effects of contextual information informing real-world actions. This result motivates the use of CI in our study and others to investigate privacy norms in realistic situations. 

In 2016, Shvartzshnaider et al. used the language of CI to survey crowdworkers' privacy expectations regarding information flow in the education domain \cite{shvartzshnaider2016learning}. Survey respondents indicated whether information flows situated in clearly defined contexts violated acceptability norms. 
The results were converted into a logic specification language which could be used to verify privacy norm consistency and identify additional acceptable information flows. 

In 2018, we designed a scalable survey method for discovering privacy norms using questions based on CI \cite{noah}. We applied the survey method to measure the acceptability of 3,840 information flows involving common connected devices for consumer homes. 
Results from 1,731 Amazon Mechanical Turk respondents informed recommendations for IoT device manufacturers, policymakers, and regulators. 

This paper adapts the survey method from our previous work \cite{noah} for a specific application: comparing privacy norms to privacy regulation.
Our use of language from regulation in CI survey questions, direct comparison of discovered privacy norms to policy compliance plans, and survey panel of special interest individuals (parents of children under age~13) distinguishes our work from previous uses of the survey method and previous CI studies in general. 

\subsection{COPPA \& Smart Toys}

Previous research has investigated Internet-connected toys and COPPA from various perspectives. Several studies have focused on identifying privacy and/or security vulnerabilities of specific smart toys \cite{valente2017security, streiff2018s, shasha2018smart}, some of which are expressly noted as COPPA violations \cite{chu2018security}. Our work uses these examples to inform the information flow descriptions included on our survey.  

Researchers have also developed methods to automate the detection of COPPA violations. 
In 2017, Zimmeck et al. automatically analyzed 9,050 mobile application privacy policies and found that only 36\% contained statements on user access, editing, and deletion rights required by COPPA \cite{zimmeck2017automated}. 
In 2018, Reyes et al. automatically analyzed 5,855 Android applications designed for children and found that a majority potentially violated COPPA \cite{reyes2018won}. 
Most violations were due to collection of personally identifiable information or other identifiers via third-party software development kits (SDKs) used by the applications, often in violation of SDK terms of service. 
These widespread violations indicate that COPPA remains insufficiently enforced.
Nevertheless, COPPA remains the primary legal foundation for state \cite{nys-truste} and federal \cite{coppa-vtech} action against IoT toy manufacturers and other technology companies for children's privacy breaches. 

Additional work has investigated parents' and children's relationships with Internet-connected toys. 
In 2015, Manches et al.~conducted observational fieldwork of children playing with Internet-connect toys and held in-school workshops to investigate parents' and children's cognizance  of how IoT toys work \cite{manches2015three}. They found that most children and caregivers were unaware of IoT toys' data collection potential, but quickly learned fundamental concepts of connected toy design when instructed. 

In 2017, McReynolds et al.~conducted interviews with parents and children to understand their mental models of and experience with Internet-connected toys \cite{mcreynolds2017toys}. Parents in this study were more aware of and concerned about IoT toy privacy than in \cite{manches2015three}, likely due to the intervening two years of negative publicity about connected toy privacy issues. The parents interviewed by McReynolds et al. provided feedback about desired privacy properties for connected toys, such as improved parental controls and recording indicators. The researchers urge ongoing enforcement of COPPA, but do not evaluate the parents' responses in light of the law.

Our work builds on past research by obtaining opinions about smart toy information collection and transfer practices from a much larger pool of parents (195 subjects). We use these data to evaluate whether privacy protections mandated by COPPA align with parents' privacy norms. 

\section{CI Survey Method}
This study adapts a CI-based survey method first presented in our previous work~\cite{noah} to evaluate whether specific requirements in privacy regulations align with user privacy norms. 
We chose this particular survey method because it is previously tested, scalable to large respondent populations, and easily adaptable to specific domains. 
The survey method works as follows, with our modifications for regulation analysis marked in italics:
\begin{enumerate}
    \item Information transfers (``flows'') are defined according to CI as sets of five parameters: subject, sender, attribute, recipient, and transmission principle (described in Section \ref{sec:CI}).
    \item We select lists of values for each of these parameters \textit{drawn from or directly relevant to a particular piece of privacy regulation.} Using these values, we generate a combinatorial number of information flow descriptions \textit{allowed or disallowed by the regulation.}
    \item Survey respondents rate the acceptability of these information flows, each of which describe a concrete data collection scenario in an understandable context.
    \item Comparing the average acceptability of flows \textit{allowed or disallowed by the regulation} indicates how well they align with respondents' privacy norms.
    \item Variations in acceptability contingent upon specific information flow parameters or respondent demographics can reveal nuances in privacy norms \textit{that may or may not be well served by the regulation.}
\end{enumerate}
The following sections provide detailed descriptions of our survey design (Sections~\ref{sec:info-flows}--\ref{sec:survey-design}), deployment (Section~\ref{sec:survey-deployment}), and results analysis (Section~\ref{sec:results-analysis})
for comparing parents' privacy norms about smart toy data collection against COPPA regulation. Many of these steps mirror those in our previous work~\cite{noah}, but we include them here with specific details from this study for the sake of replicability.

\subsection{Generating Smart Toy Information Flows}
\label{sec:info-flows}

We first selected CI information flow parameters (Table~\ref{fig:elements}) involving smart toys and specific data collection requirements from COPPA.
We then programmatically generated information flow descriptions from all possible combinations of the selected CI parameters. 

We next discarded certain information flow descriptions with
unrealistic sender/attribute pairs, 
such as a toy speaker (sender) recording a child's heart rate (attribute).
Unrealistic sender/attribute pairs were identified at the authors' discretion based on whether each toy could reasonably be expected to have access to each type of data during normal use. This decision was informed by smart toy products currently available on the market.
The use of exclusions to remove unrealistic information flows is a core part of the CI survey method~\cite{noah} for reducing the total number of questions and the corresponding cost of running the survey. 
This process resulted in 1056 total information flow descriptions for use in CI survey questions (Section~\ref{sec:survey-design}).

The degree to which these flows are rated as acceptable or unacceptable by survey respondents indicate agreement or disagreement between COPPA and parents' privacy norms. This rest of this section describes how we selected values for each information flow parameter in detail. 

\begin{table*}[t]
\renewcommand{\arraystretch}{1.1}
\centering
\small

\begin{tabular}{p{5.1cm}|p{11.6cm}}
\bottomrule
\textbf{Sender} & \textbf{Transmission Principle}\\
\hline
a smart speaker/baby monitor & \textit{COPPA Compliance Plan Steps 2-3}\\
a smart watch & if its privacy policy permits it \\
a toy walkie-talkie & if its owner is directly notified before the information was collected\\
a smart doll &  \\
a toy robot & \textit{COPPA Compliance Plan Step 4} \\
& if its owner has given verifiable consent\\
\textbf{Recipient} & if its owner has given verifiable consent before the information was collected\\
\cline{1-1}
its manufacturer &  \\
a third-party service provider & \textit{COPPA Compliance Plan Step 5}\\
& if its owner can at any time revoke their consent, review or delete the information collected \\
\textbf{Subject \& Attribute}& \\
\cline{1-1}
its owner's child's heart rate & \textit{COPPA Compliance Plan Step 6}\\
its owner's child's frequently & if it implements reasonable procedures to protect the information collected\\
\hspace{1em} asked questions & if the information is kept confidential\\
the times its owner's child is home & if the information is kept secure\\
its owner's child's frequently & if the information is stored for as long as is reasonably necessary for the purpose  \\
\hspace{1em} traveled routes & \hspace{1em} for which it was collected\\
the times it is used & if the information is deleted  \\
its owner's child's location &  \\
its owner's child's sleeping habits & \textit{COPPA Exclusions}\\
its owner's child's call history & if the information is used to protect a child's safety\\
audio of its owner's child & if the information is used to provide support for internal operations of the device\\
its owner's child's emergency contacts & if the information is used to maintain or analyze the function of the device\\
video of its owner's child & if the information is used to serve contextual ads \\
its owner's child's birthday & \\
& \textit{Other} \\
& if it complies with the Children's Online Privacy Protection Rule\\
& \textit{null}\\
\toprule
\end{tabular}
\caption{Contextual integrity parameter values selected for information flow generation.  The \textit{null} transmission principle is an important control included to generate information flows with no explicit conditions. The transmission principles were derived from the FTC's Six Step Compliance Plan for COPPA \cite{federalsteptrade}. 
}
\label{fig:elements}
\end{table*}

\paragraph{Transmission Principles from COPPA.}
\label{sec:tp}
We used the Federal Trade Commission's Six Step Compliance Plan for COPPA \cite{federalsteptrade} to identify transmission principles. 
Some of these transmission principles match those in our previous work \cite{noah}, facilitating results comparison. 

We converted steps 2--4 of the Compliance Plan into four transmission principles regarding consent, notification, and privacy policy compliance (Table~\ref{fig:elements}).
COPPA dictates that parents must receive direct notice and provide verifiable consent before information about children is collected. Operators covered by COPPA must also post a privacy policy that describes what information will be collected and how it will be used. 
Our corresponding transmission principles allow us to test whether these requirements actually increase the acceptability of data collection from and about children.

The fifth step of the Compliance Plan concerns ``parents' ongoing rights with respect to personal information collected from their kids'' \cite{federalsteptrade}. Operators must allow parents to review collected information, revoke their consent, or delete collected information. We translated this requirement into the transmission principle ``if its owner can at any time revoke their consent, review or delete the information collected.''

The sixth step of the Compliance Plan concerns operators' responsibility to implement ``reasonable procedures to protect the security of kids' personal information''~\cite{federalsteptrade} and to only release children's information to third party service providers who can do likewise. 
We translated this step into five transmission principles involving confidentiality, security, storage and deletion practices (Table~\ref{fig:elements}).

The Compliance Plan also lists a set of exclusions to COPPA. We converted the exclusions that were most applicable to Internet-connected  children's devices into four transmission principles (Table~\ref{fig:elements}). 
We also added the transmission principle ``if it complies with the Children's Online Privacy Protection Rule'' to test parents' trust and awareness of COPPA itself.

Importantly, we also included the \textit{null} transmission principle to create control information flows with no COPPA-based criteria. Comparing the acceptability of flows with the \textit{null} transmission principle against equivalent flows with COPPA-based transmission principles allows us to determine whether the COPPA conditions are relevant to parents' privacy norms. 

\paragraph{Smart Toy Senders.}
The senders included in our survey represent five categories of children's IoT devices: a smart speaker/baby monitor, a smart watch, a toy walkie-talkie, a smart doll, and a toy robot. 
We chose these senders by searching for children's Internet-connected devices mentioned in recent press articles \cite{finkle_bartz_patnaik_2015,techcrunch_2017,nienaber;_2017,peachman_2017, moon_2017, reuters_1}, academic papers \cite{mahmoud2017towards, euhub_2017}, blogs \cite{iotlist, intsoc, consumerist}, IoT-specific websites \cite{privacyincluded, which, kidsafe}, and merchants such as Toys ``R'' Us and Amazon. 
All of the selected senders are devices that are reasonably ``directed towards children'' \cite{federalsteptrade,federaltradecoppa} in order to ensure that they are covered by COPPA. 
We excluded devices such as smart thermometers or other smart home devices that might collect information about children but are not directly targeted at children.

It is important to note that the selected devices do not represent the full breadth of smart toy products. However, information flow descriptions involving specific devices or device categories evoke more richly varied privacy norms from survey respondents than flows describing a generic ``smart toy.'' This is supported by existing interview data~\cite{zheng2018smart} noting that IoT device owners often have very different privacy opinions of specific entities than of their generic exemplars (e.g., the ``Seattle government'' versus ``government''). 

\paragraph{Information Attributes.}
We reviewed academic research~\cite{mahmoud2017towards}, online privacy websites \cite{privacyincluded}, toy descriptions~\cite{kidswatch}, and privacy policies \cite{pilab,toytalk} to compile a list of information attributes collected by the toys in our sender list. 
The final selected attributes include heart rate, frequently asked questions, the times the subject is home, frequently traveled routes, the times the device is used, location, sleeping habits, call history, audio of the subject, emergency contacts, video of the subject, and birthday. 
These attributes cover a variety of personally identifiable or otherwise sensitive information with specific handling practices mandated by COPPA.

\paragraph{First- and Third-party Recipients.}
We included device manufacturers and third-party service providers as recipient parameters. This allowed us to examine variations in privacy between first and third parties while limiting the total number of information flows and the corresponding cost of running the survey. 

\paragraph{Children as Information Subjects.}
\label{sec:subject}
The only subject parameter included in the survey is ``its owner's child.'' 
This wording emphasizes that the child is not the owner of the device and acknowledges the parental role in ensuring children's privacy. It also accounts for devices that may not be used directly or exclusively by the child (e.g., a baby monitor). 
We indicated in the survey overview that respondents should think about their own children's information when interpreting this subject. 

\subsection{Survey Design}
\label{sec:survey-design}

We created and hosted the survey on the Qualtrics platform~\cite{qualtrics}. The survey was split into six sections: consent, demographic questions~I, overview, contextual integrity questions, awareness questions, and demographic questions~II. 
This section provides details about each section.
The survey did not mention COPPA, privacy, security, nor any potential negative effects of smart toy information flows prior to the contextual integrity questions to prevent priming and framing effects.

\paragraph{Consent.}
Respondents were initially presented with a consent form approved by our university's Institutional Review Board. Respondents who did not consent to the form were not allowed to proceed with the study. 

\paragraph{Demographic Questions I.}
The first set of demographic questions asked respondents for the ages of their children under 13.  We chose this age limit because COPPA only applies to data collection from children under~13. We randomly selected one of the ages for each respondent, $n$, which was piped to the survey overview.

\paragraph{Overview.}
Respondents were then presented with a survey overview containing a brief description of Internet-connected devices and instructions for the contextual integrity questions (Appendix~A). 
This overview also explained how respondents should interpret the recurring phrase ``its owner's child,'' and instructed them to keep their $n$-year-old child in mind while taking the survey (where $n$ was selected for each respondent from their responses to the demographics questions I). 

\paragraph{Contextual Integrity Questions.}
The core of the survey consisted of 32 blocks of questions querying the acceptability of our generated information flows (Section~\ref{sec:info-flows}). 
Each question block contained 33 information flows with the same sender, same attribute, varying recipients, and varying transmission principles. For example, one block contained all information flows with the sender ``a smart doll'' and the attribute ``the times it is used.'' Each question block also included one attention check question.

Each respondent was randomly assigned to a single question block. Answering questions about flows with the same sender and attribute reduced cognitive fatigue and ensured independence across recipients and transmission principles. 

\begin{figure*}[t]
\centering
\fbox{\includegraphics[width=0.8\textwidth]{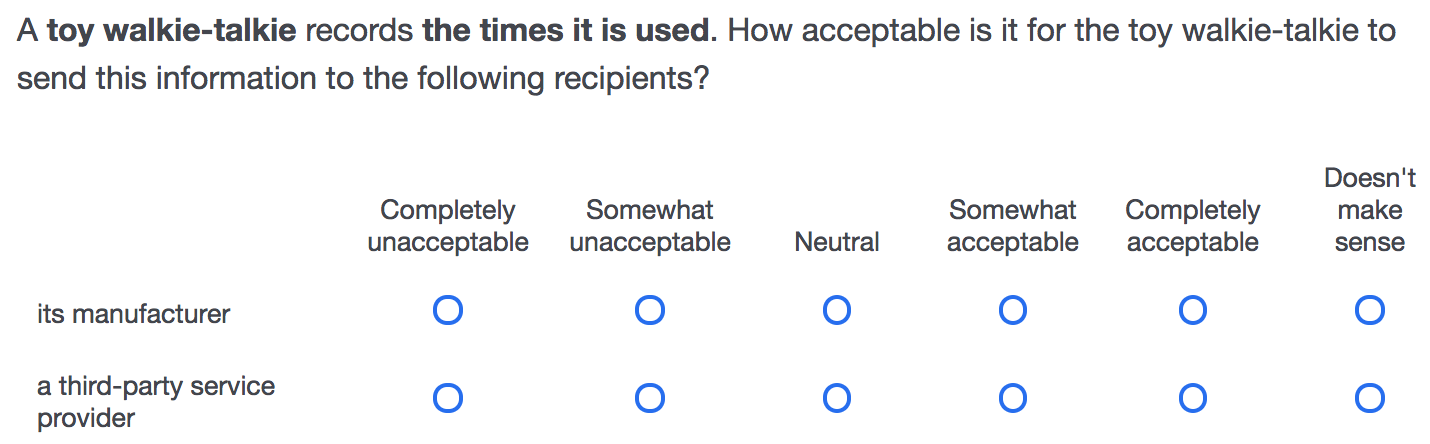}}
\caption{Example CI question matrix with information flows to different recipients and the \textit{null} control transmission principle.}
\label{fig:nulltp-question}
\end{figure*}

\begin{figure*}[t]
\centering
\fbox{\includegraphics[width=0.8\textwidth]{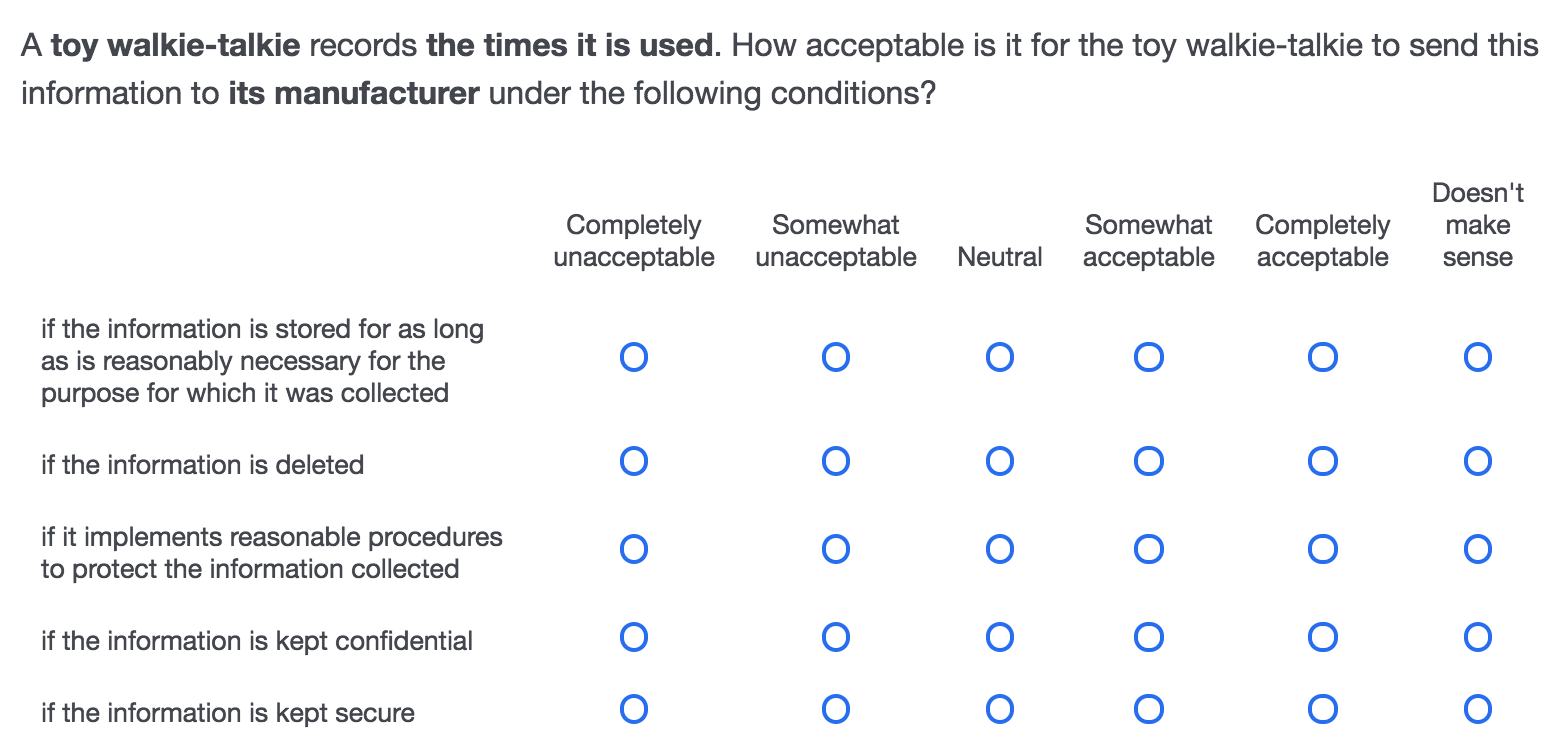}}
\caption{Example CI question matrix with information flows to a fixed recipient and varying transmission principles.}
\label{fig:tp-question}
\end{figure*}

The information flows in each block were divided into matrices of individual Likert scale multiple choice questions. The first matrix in each block contained questions about information flows to different recipients with the \textit{null} transmission principle (Figure~\ref{fig:nulltp-question}). 
The remaining matrices each contained questions about information flows to a specific recipient with varying transmission principles (Figure~\ref{fig:tp-question}). 
The order of the information flows in each block was randomized for each respondent. 

Each individual multiple choice question in the matrices asked respondents to rate the acceptability of a single information flow on a scale  
of five Likert items: Completely Acceptable (2), Somewhat Acceptable (1), Neutral (0), Somewhat Unacceptable (-1), Completely Unacceptable (-2).
We also included the option ``Doesn't Make Sense'' to allow respondents to indicate if they didn't understand the information flow.

\paragraph{Awareness Questions.}
Respondents then answered questions about their general technological familiarity and Internet use, ownership of Internet-connected devices, ownership of  children's Internet-connected devices, and previous knowledge of COPPA.

\paragraph{Demographic Questions II.}
Finally, respondents answered standard demographic questions from the United States Census. This allowed us to check the representativeness of our sample (Appendix~B, Section~\ref{sec:representativeness}) and account for demographic variables in our analysis. 

\subsection{Survey Deployment}
\label{sec:survey-deployment}

We tested the survey on UserBob~\cite{userbob} once during the survey design process and again immediately prior to deployment. 
UserBob is a usability testing service for obtaining video screen capture of users interacting with a website while recording audio feedback. 
Each survey test involved creating a UserBob task with a link to the survey, brief instructions for users,\footnote{UserBob task instructions: ``This is a survey that will be given to a group of parents with children younger than 13. Take the survey, pretending you have one or more children younger than 13. Record your thoughts on the user interface and whether the questions do/don't make sense.''} 
and settings to recruit 4 users to take the survey for 7 minutes each. UserBob automatically recruited users through Amazon Mechanical Turk at a cost of $\$1$ per user per minute. 
The resulting video and audio recordings of users interacting with the survey informed changes to our survey design. In particular, we reduced the number of questions per block and increased the number of pages over which the questions were presented. This reduced the amount of scrolling necessary to complete the survey and improved engagement. 
This practice of using pre-deployment ``cognitive interviews'' to test and debug survey design is common in survey research~\cite{sudman1996thinking}.
UserBob responses were not included the final results.

We used Cint~\cite{cint}, an insights exchange platform,
to deploy our survey to a panel of 296 adult parents of children under the age of 13 in the United States. 
We selected 
respondents with children younger than~13 because COPPA applies to ``operators of websites or online services directed to children under~13''~\cite{federaltradecoppa}. 
Our surveyed population therefore consisted entirely of individuals affected by COPPA. 
We chose not to set a minimum age for respondents' children, because there is a lack of readily available information on the minimum age of use of Internet-connected children's devices. 
While certain manufacturers list recommended minimum ages for their connected toys and devices, this was not the case for the majority of the devices we considered. Additionally, many devices such as wearable trackers, water bottles, baby monitors, are targeted towards very young children.
Lastly, not restricting the minimum age allowed us to relax the demographic requirements for survey deployment. 

Respondents were paid \$3 for valid responses where the attention check question was answered correctly.
Each respondent was only allowed to answer the survey once. The survey responses were collected over an 18 hour time frame.
We chose Cint to deploy our survey instead of Amazon Mechanical Turk, because Cint allowed us to directly target a specific panel of respondents (as in Zyskowski et al.~\cite{zyskowski2015accessible}) without requiring a preliminary screening questionnaire to identify parents~\cite{schleider2015using}.
\subsection{Response Analysis}
\label{sec:results-analysis}

We began with 296 responses. We removed the responses from 8 respondents who did not consent to the survey (none of their information was recorded) as well as those 
from 85 respondents who did not correctly answer the attention check question. We removed 2 responses in which over 50\% of the information flows were characterized as ``Doesn't make sense.'' We also removed 2 responses where not all information flow questions were answered. Finally, we removed 1 response where the respondent self-reported over 10 children and 3 responses that were completed in less than 2 minutes. This resulted in a final set of 195 responses with an average of 6 responses per information flow (standard deviation~1.4).

The responses to all contextual integrity questions (Section~\ref{sec:survey-design}) were on a Likert scale with the following Likert items: ``Completely acceptable''~(2), ``Somewhat acceptable''~(1),  ``Neutral''~(0), ``Somewhat unacceptable''~(-1), and ``Completely unacceptable''~(-2). We call this value the ``acceptability score'' of each information flow for each respondent. 

\begin{figure*}[tp]
\centering
\includegraphics[width=\textwidth]{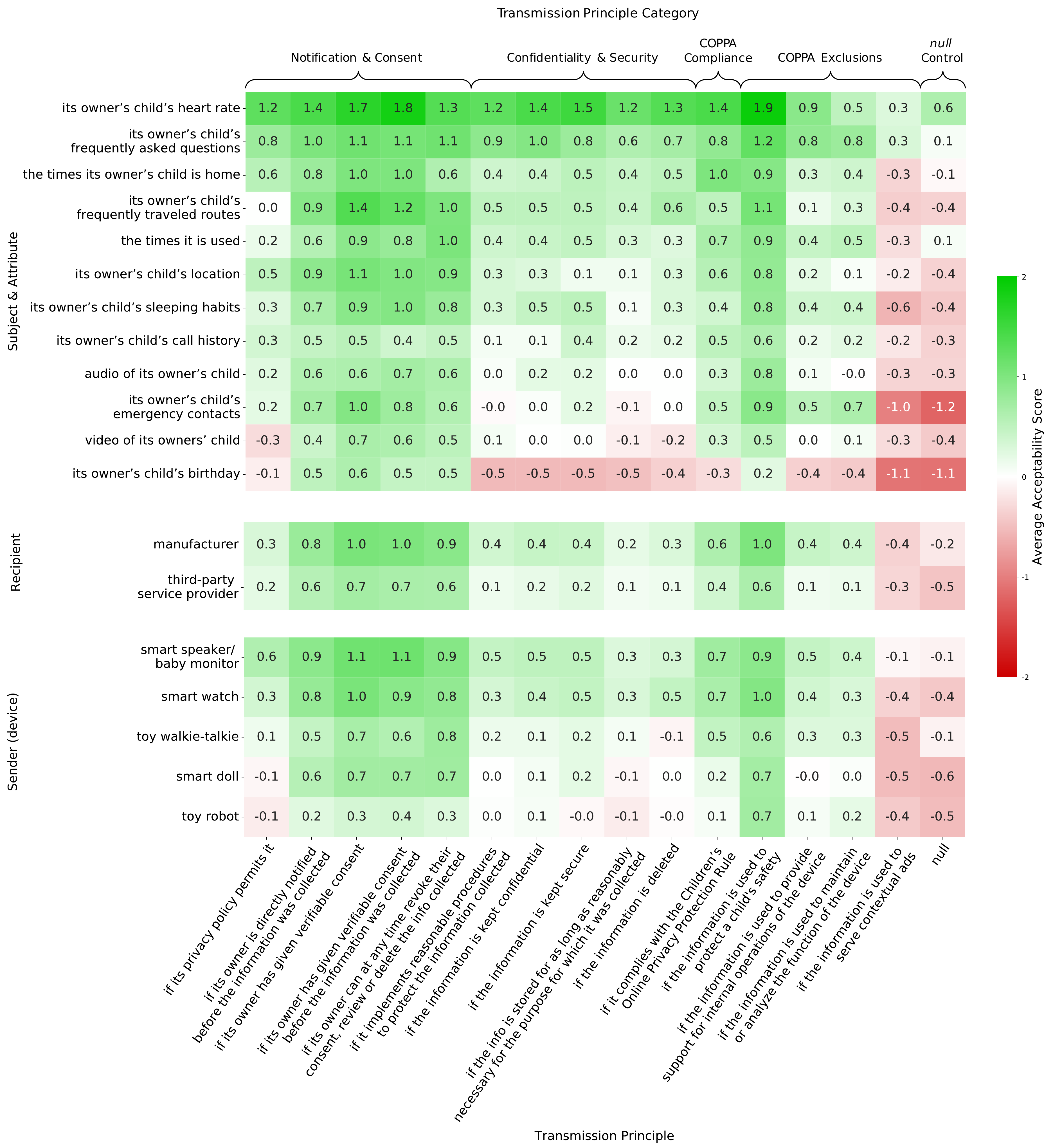}
\caption{Average acceptability scores of information flows grouped by COPPA-derived transmission principles and attributes, recipients, or senders. Scores range from $-2$ (completely unacceptable) to $2$ (completely acceptable). }
\label{fig:flows-results}
\end{figure*}

\begin{figure*}[tp]
\centering
\includegraphics[width=0.99\textwidth]{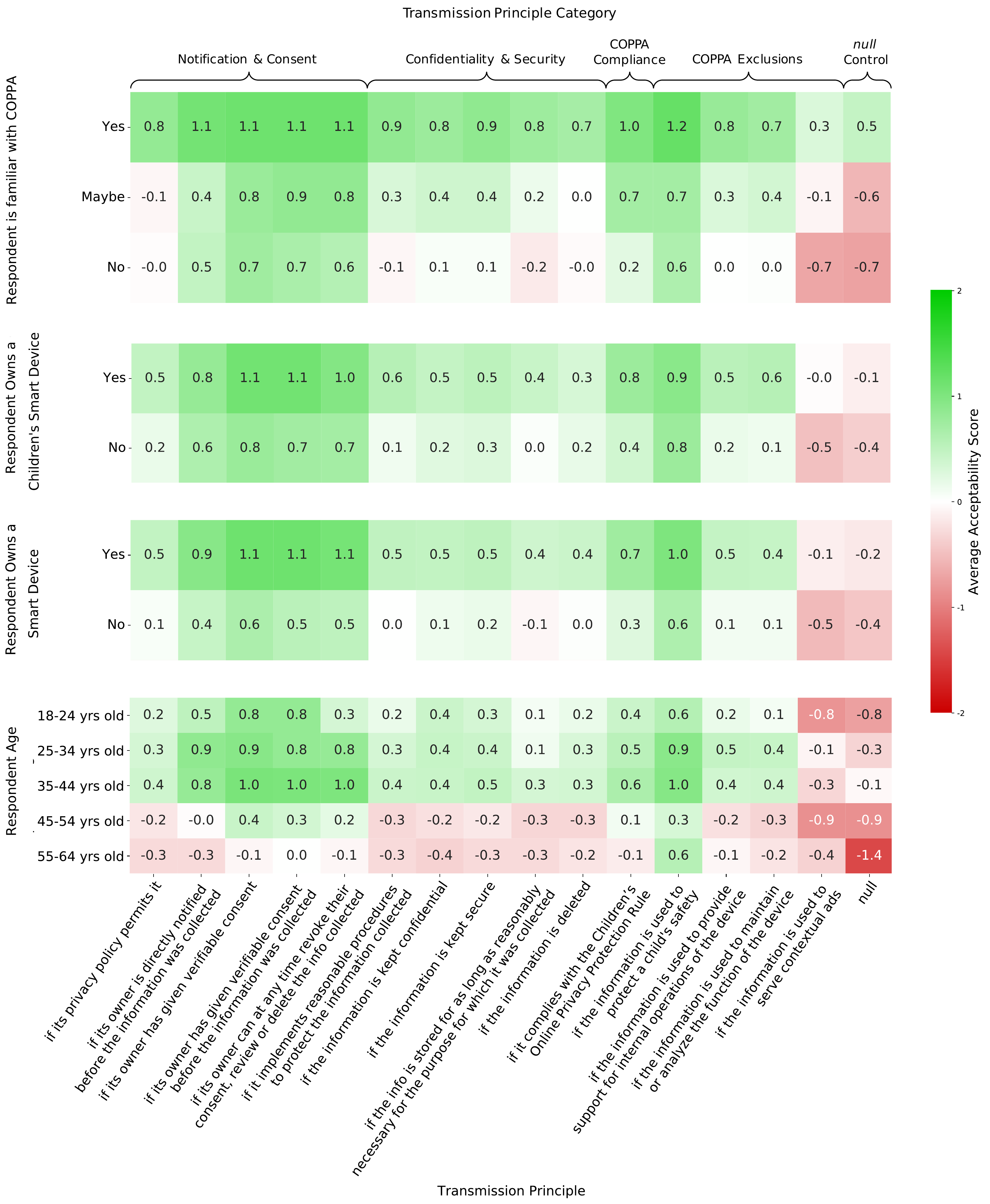}
\caption{Average acceptability scores of information flows grouped by COPPA-derived transmission principles and respondent ages, familiarity with COPPA, or ownership of smart devices. Scores range from $-2$ (completely unacceptable) to $2$ (completely acceptable)}
\label{fig:demo-results}
\end{figure*}

In order to generalize privacy norms beyond individual respondents and information flows, we averaged the acceptability scores of flows grouped by CI parameters or respondent demographics.  
For example, we averaged the acceptability scores of all information flows with the recipient ``its manufacturer'' and the transmission principle ``if the information is deleted'' in order to quantify the pairwise effects of these two parameters on privacy norms. We then plotted these pairwise average acceptability scores as heatmaps to visualize how individual CI parameters or respondent demographic factors affect the overall alignment of information flows with privacy norms (Figures~\ref{fig:flows-results}~\&~\ref{fig:demo-results}).

We statistically compared the effects of different COPPA provisions (Sections \ref{sec:results-coppa-overall}--\ref{sec:results-consent-vs-security}) by averaging the acceptability scores of all information flows grouped by transmission principles.  
For example, one group contained the average score given by each of the 195 respondents to information flows with non-\textit{null} transmission principles, while a second group contained the average score given by each respondent to information flows with the \textit{null} transmission principle. 
We then applied the Wilcoxon signed-rank test to find the likelihood that these two groups of scores come from the same distribution. We performed three such tests with different transmission principle groups and set the threshold for significance to $p = 0.05/3 = 0.016$ to account for the Bonferroni multiple-testing correction.

We statistically compared the effects of smart device awareness, COPPA familiarity, and demographic factors (Sections~\mbox{\ref{sec:results-demo-coppa}--\ref{sec:results-demo-income-edu}}) by averaging the acceptability scores of all information flows grouped by respondent category of interest. 
For example, one group contained the average score given by each respondent who owned a smart device across all answered CI questions, while the second set contained the average score given by each respondent who did not own a smart device.
We then applied the Wilcoxon signed-rank test to find the likelihood that these two groups of scores come from the same distribution. We performed five such tests with groupings based on COPPA familiarity, age, smart device ownership, education, and income and set the threshold for significance to $p = 0.05/5 = 0.01$ to account for the Bonferroni multiple-testing correction. 

\section{Results}
Overall, surveyed parents view information flows meeting COPPA data collection guidelines as acceptable while viewing equivalent information flows without COPPA criteria as unacceptable (Figures~\ref{fig:flows-results}~\&~\ref{fig:demo-results}). This supports the conclusion that COPPA-mandated information handling practices generally align with parents' privacy norms. In this section, we elaborate on this finding and explore additional trends in our survey responses to further compare COPPA to parents' privacy norms regarding children's smart toys.

\subsection{COPPA Data Collection Requirements\\Align with Parents' Privacy Norms}
\label{sec:results-coppa-overall}
COPPA requirements were incorporated in the survey as information flow transmission principles derived from the FTC's Six-Step Compliance Plan for COPPA~\cite{federalsteptrade} (Section~\ref{sec:info-flows}). The average acceptability scores of information flows explicitly obeying these requirements are mostly non-negative (Figures~\ref{fig:flows-results}~\&~\ref{fig:demo-results}). This indicates that most surveyed parents consider these flows as ``completely acceptable'' or ``somewhat acceptable.'' In comparison, the average acceptability scores of information flows with the control \textit{null} transmission principle are mostly negative (Figures~\ref{fig:flows-results}~\&~\ref{fig:demo-results}), indicating that most surveyed parents consider these flows without COPPA criteria as ``completely unacceptable'' or ``somewhat unacceptable.'' 

This difference between information flows with no explicit conditions versus flows with COPPA requirements holds regardless of information sender, recipient, attribute, or parents' demographics (apart from a few specific exceptions which we discuss below). On average, information flows with COPPA-derived transmission principles are 0.73 Likert-scale points
more acceptable than their \textit{null} transmission principle counterparts ($p < 0.001$). 

Our research provides the first quantitative evidence that COPPA guidelines generally match parents' privacy norms for Internet-connected toys. This indicates that regulation can mandate meaningful transmission principles for information flows and supports further creation and fine-tuning of regulation to keep Internet data collection within the bounds of consumer privacy preferences. 

\subsection{Parents View Data Collection for\\Contextual Advertising as Unacceptable}
Information flows with the transmission principle ``if the information is used to serve contextual ads'' have negative average acceptability scores across almost all senders, recipients, and attributes (Figure~\ref{fig:flows-results}). 
Unlike all other information flows on our survey with non-\textit{null} transmission principles, these flows are actually prohibited by COPPA. 
The ``contextual ads'' transmission principle is a ``limited exception to COPPA's verifiable parental consent requirement'' as listed in the COPPA Compliance Plan~\cite{federalsteptrade}. This exception only applies to the collection of persistent identifiers (such as cookies, usernames, or user IDs) and not to any of the attributes included on our survey.
Our respondents generally agree that collecting the attributes on our survey for contextual (targeted) advertising would be unacceptable, providing further support for COPPA's alignment with parents' norms.

This result indicates that the CI survey technique can detect regulatory provisions that reduce alignment with privacy norms, essential for future applications of the method (Section~\ref{sec:policy-applications}). It also provides evidence that the mere presence of a transmission principle doesn't necessarily improve the acceptability of information flows. 

This result relates to existing work about opinions of data collection for advertising. Zheng et al.~\cite{zheng2018smart} interviewed owners of non-toy Internet-connected home devices and found mixed opinions of targeted advertising with data from these devices depending on the perceived benefit to the user. Combined with our results, this suggests that parents do not believe that relaxing COPPA to allow contextual advertising from more types of children's toy data would have enough benefit to outweigh privacy concerns. 

\subsection{Parents View Children's Birthdays as\\Especially Private}
Information flows including the subject and attribute ``its owner's child's birthday'' are an exception to the trend described in Section~\ref{sec:results-coppa-overall}.  The average acceptability scores of information flows with this attribute and 10 of the 15 COPPA-derived transmission principles are negative (Figure~\ref{fig:flows-results}). This discrepancy could be attributed to the relatively small number of parents (11 parents or $5.6\%$ of total respondents) who were asked to score flows with this attribute. Parents may also view their children's birthdays as more personal than the other surveyed attributes or as less necessary for some of the surveyed transmission principles (such as ``to maintain or analyze the function of the device''). Follow-up qualitative studies could focus on specific attributes, such as children's birthdays, to understand parents' rationales behind corresponding privacy norms. 

\subsection{Notification \& Consent Versus\\Confidentiality \& Security}
\label{sec:results-consent-vs-security}

Our results also provide insights into the relative importance of different sections within COPPA to parents' privacy norms. This could help regulators prioritize certain forms of non-compliant information collection for legal action. 

Our COPPA-derived transmission principles can be divided into categories based on their topic and the section of the COPPA Compliance Plan~\cite{federalsteptrade} from which they were drawn (Section~\ref{sec:info-flows}). One category consists of transmission principles from the Compliance Plan steps 2--5 regarding notification and consent  (Table~\ref{fig:elements}). These transmission principles involve device privacy policies, the collection of verified consent, and the ability to revoke consent or review collected information. 
Another category consists of transmission principles from the Compliance Plan step 6 regarding information confidentiality and security (Table~\ref{fig:elements}). These transmission principles involve reasonable data protection, confidential and secure storage, and limited information lifetime. 

Across all senders, attributes, and recipients, information flows with transmission principles in the notification/consent category have significantly higher acceptability scores than flows with transmission principles in the confidentiality/security category by an average of $0.43$ Likert scale points ($p < 0.001$) 
(Figure~\ref{fig:flows-results}). 
One notable exception to this trend is the transmission principle ``if its privacy policy permits it.'' The acceptability scores for this transmission principle are an average of $0.53$ Likert points
lower than for others in the notification/consent category ($p < 0.001$). We suspect this reflects the general distrust of privacy policies evidenced in previous research \cite{turow2001privacy}. Privacy policies are typically dense, lengthy, and difficult to interpret even for experts \cite{reidenberg2015disagreeable}. It therefore makes sense that parents would not view the disclosure of information collection in privacy policies as acceptable as other notification methods. 

The greater acceptability of information flows with notification or consent criteria versus flows with confidentiality or security criteria corroborates previous research using the CI survey method to discover privacy norms of non-toy consumer IoT devices \cite{noah}. This provides longitudinal data indicating that users of Internet-connected products continue to prioritize consent over built-in security when reasoning about the appropriateness of information collection practices. This motivates continued work to improve the state of notification and consent mechanisms for Internet data collection. The most prevalent mechanisms, privacy policies and mobile application permissions, are widely understood to be ineffective for informing users or providing meaningful privacy control options \cite{solove2012}. As policies change to nuance the definitions of informed consent to include ideas of intelligibility, transparency and active opt-in, among others, it is important to continue to study and evaluate consumer's privacy expectations regarding consent.

\subsection{COPPA Compliance and Familiarity\\Increase Data Collection Acceptability}
\label{sec:results-demo-coppa}

Information flows with the transmission principle ``if it complies with the Children's Online Privacy Protection Rule'' received a positive average acceptability score of 0.49 across all senders, recipients, and attributes. As expected, flows with this transmission principle were rated as more acceptable by the 67\% of respondents familiar with COPPA than by the 33\% of respondents unfamiliar with the rule. 

Furthermore, respondents who indicated that they were familiar with COPPA rated all information flows $0.75$ Likert points more acceptable on average than respondents who were not familiar with the rule 
($p < 0.001$)
(Figure~\ref{fig:demo-results}). 

In both cases, stated compliance and/or familiarity with COPPA may increase parents' acceptance of smart toy data collection by reassuring them that their children's privacy is protected by regulation. However, this may be a false sense of security, as COPPA guidelines are relatively broad and COPPA violations are likely widespread in practice (Section~\ref{sec:policy-insights})~\cite{chu2018security, reyes2018won}.

\subsection{Younger Parents are More Accepting of\\Smart Toy Data Collection}
\label{sec:results-demo-age}

Parents younger than 45 gave an average acceptability score of $0.48$ to all rated flows, following the trend discussed in Section~\ref{sec:results-coppa-overall} (Figure~\ref{fig:demo-results}). In comparison, parents 45 years and older gave an average acceptability score of $-0.17$ to all rated flows. This difference in the acceptability scores of these two groups is significant ($ p < 0.01$).
Nevertheless, context still matters, as information flows specifically ``to protect a child's safety'' are viewed as generally acceptable to all surveyed parents regardless of age. 

Previous work indicates that young American adults are more aware of online privacy risks and more likely to take steps to protect their privacy online than older adults \cite{rainie2016state}. Future studies could investigate why this awareness of online privacy risks makes younger parents more accepting of smart toy data collection. 

\subsection{Parents Who Own Smart Devices are\\More Accepting of Data Collection}
\label{sec:results-demo-smartdevice}

Parents who own generic smart devices or children's smart devices were more accepting of information flows than respondents who do not own these devices on average, but the difference in scores (0.34 Likert scale points) between these two groups is not significant ($p = 0.12$).

Nevertheless, this difference corroborates previous work using the CI survey method, in which owners of non-toy consumer IoT devices were found to be more accepting of information flows from these devices than non-owners \cite{noah}. This difference likely reflects a self-selection bias, in which those more uncomfortable with Internet data collection are less likely to purchase Internet-connected toys or other devices. However, the small effect size in both this study and the previous work may be due to parents purchasing smart toys unaware of their data collection potential \cite{manches2015three} or willing to trade-off privacy concerns for other benefits provided by the products~\cite{zheng2018smart}.

\subsection{Education \& Income have Little Effect on\\Parents' Smart Toy Privacy Norms}
\label{sec:results-demo-income-edu}
Parents' education and income did not have significant effects on acceptability scores. Parents earning more than $\$100,000$ per year gave an average acceptability score of $0.46$ to all rated flows, not significantly different from the  average score of $0.37$ from parents earning less ($p = 0.77$). Similarly, parents with at least some college education gave an average acceptability score of $0.37$, not significantly different from the $0.33$ average score of parents with a high school diploma or less ($p = 0.58$). This is perhaps unexpected given previous work indicating that parents with more resources are more likely to engage with children on privacy issues~\cite{redmiles2018} and is a topic for follow-up research. 

\section{Limitations}
\label{sec:limitations}
Our results must be considered in the context of the following limitations.

\subsection{Privacy Attitudes Versus Behaviors}
\label{sec:privacy-paradox}
Individuals often self-report greater privacy awareness and concerns than reflected in actual privacy-related behaviors~\cite{acquisti2015privacy, kokolakis2017privacy}. This ``privacy paradox'' is well-documented and poses a challenge for researchers. 
The CI survey method is vulnerable to privacy paradox effects. However, 
there is a reasonable argument that privacy regulation should prioritize the expressed norms of users (measured by the survey instrument) over norms evidenced through behaviors, which are influenced by external factors (such as confusing user interfaces) that could be affected by the regulation. For example, it is often difficult for consumers to determine the data collection practices of IoT devices, including Internet-connected children's toys, due to poor company disclosure practices~\cite{reidenberg2015disagreeable} and limited auditing by third parties. Just because many parents purchase smart toys does not mean that they approve of the toys' data collection practices and wouldn't support new regulation to improve privacy. 

\subsection{Respondent Representativeness}
\label{sec:representativeness}
The self-reported demographics of our respondents (Appendix~B) indicate that the sample, while diverse, is non-representative in ways that may influence measured privacy norms.

Females and high-income individuals are notably overrepresented in our sample compared to the United States population. The literature on gender differences in online privacy concerns suggests that women may generally perceive more privacy risks online than men \cite{bartel1999investigation, fogel2009internet, youn2008gender}, but some studies contradict this conclusion, reporting no significant gender effect~\cite{yao2007predicting}. The effect of income on online privacy concerns is similarly unsettled, with some reporting that high-income individuals are less concerned about privacy~\cite{o2001analysis, madden2017privacy}, others reporting that high-income individuals are more likely to engage in privacy-preserving behaviors~\cite{redmiles2018}, and still others finding no significant income effect~\cite{zhang2002characteristics}. 

Limiting our survey to parents also ignores the opinions of other parties, including school and daycare teachers and extended family members, who also purchase Internet-connected toys for children but may have different privacy norms. 
These individuals are also affected by COPPA and have legitimate justification for their opinions and interests to be reflected in children's privacy regulation. 
Likewise, we did not ask whether our respondents were members of communities that may have less common privacy norms, but our respondent panel, drawn from across the United States, certainly missed smaller demographics.

Finally, our respondent panel consisted entirely of parents living in the United States, as COPPA only applies to products sold in the U.S. These respondents are therefore influenced by American attitudes toward privacy, which may vary from those of parents in other countries. We hope that future work will apply the CI survey method used in this paper to evaluate the alignment between privacy norms and privacy regulation in non-U.S.~contexts.

\subsection{Goals of Privacy Regulation}
Our use of CI surveys to evaluate privacy regulation assumes that the underlying value of such regulation is to better align data collection practices with privacy norms. This makes an implicit normative argument about the purpose of privacy regulation, which does not necessarily hold, especially for the norms of majority populations. For example, privacy regulation may seek to protect minority or otherwise vulnerable populations. In these cases, surveys of all individuals affected by the regulation may reflect a majority view that does not value the norms or appreciate the situation of the target population.  CI surveys could still be applied in these contexts, but care would need to be taken to identify and recruit respondents from populations differentially affected by the regulation in order to uncover discrepancies between the regulation and the norms of these groups. 

Additionally, 
some regulation may be created with the goal of changing existing norms. In these cases, the CI survey method will indicate that the regulation does not match current privacy expectations upon enactment. However, CI surveys would still be useful for conducting longitudinal measurements to track whether the regulation has the desired effect on privacy norms over time.

\section{Discussion \& Future Work}
\label{sec:discussion}

We would like this study to serve as a template for future work using contextual integrity to analyze current or pending privacy regulation for policy or systems design insights. This section discusses our COPPA findings and presents suggestions for future applications of our method by policymakers, device manufacturers, and researchers. 

\subsection{COPPA Insights \& Concerns}
\label{sec:policy-insights}
Previous research indicates that parents actively manage the information about their children on social media platforms to avoid oversharing~\cite{ammari2015managing}, and that owners of IoT home appliances view most data collection by these devices as inherently unacceptable~\cite{noah}. We expected that these domains would overlap, resulting in skepticism of smart toy data collection that even the restrictions in COPPA could not ameliorate. Surprisingly, it seems that the COPPA criteria assuaged parents' privacy concerns on average. 

While we are encouraged that COPPA generally aligns with parents' privacy expectations, we are also concerned that the existence of COPPA may give parents an unreasonable expectation that their children's data is protected, especially since parents familiar with COPPA were less critical of smart toy information flows. In fact, several online services and Internet-connected toys have been found to violate COPPA~\cite{chu2018security, reyes2018won}, and many more non-compliant toys are likely available for purchase. Additionally, the information collection guidelines in COPPA are relatively broad, leaving room for technical implementations that adhere to the letter of the law but still compromise children's privacy.  This motivates continued work by regulators and researchers to identify toys that place children's privacy at risk, as well as healthy skepticism by parents before purchasing any particular toy.

As an additional policy insight, variations in information flow acceptability across recipients\footnote{Information flows to first-party manufacturers have higher average acceptability scores than flows to third-party service providers (Figure~\ref{fig:flows-results}).} corroborate previous work~\cite{zheng2018smart} indicating that privacy norms are deeply contingent on the perception of entities that collect online data. COPPA distinguishes between first- and third-parties, but does not further categorize data recipients. This increases the flexibility of the law, but raises the potential that some recipients, which may viewed as completely unacceptable by privacy norms, could still legally get access to children's data. This suggests that incorporating a more contextual framing of entities could improve the ability of future regulation to prevent unwanted data collection practices. 

\subsection{Further Policy Analysis Applications}
\label{sec:policy-applications}
The CI survey method is not limited to COPPA. We would like to see the results of follow-up studies focusing on different regulation, such as the Health Insurance Portability and Accountability Act (HIPAA), the Family Educational Rights and Privacy Act (FERPA), the National Cybersecurity Protection Advancement Act, the European General Data Protection Regulation (GDPR), and others from the local to international level, to see if their requirements result in similarly acceptable information flows for members of their target populations. As most privacy regulation encompasses information transfer or exchange, the theory of contextual integrity is an appropriate framework for this research. 
Further studies would also allow cross-regulatory analysis to find common factors that affect alignment with privacy norms.

The CI survey method could also be incorporated into the policymaking process.
Policy formulation and resource allocation could be guided by surveying a wide-variety of information flows allowed under current regulation and identifying egregious or unexpected norm violations that require attention.
Policymakers could test whether previous regulatory approaches will be applicable to new innovations by conducting surveys with CI parameters describing new technologies and existing regulation (e.g., smart toys and COPPA prior to the 2017 inclusion of IoT devices~\cite{futureical}).
Policymakers could also perform A/B tests of policy drafts with different stipulations and/or language by conducting multiple parallel surveys with varying CI parameters.
These and other use cases would improve quantitative rigor in data-driven policy development and facilitate the design of regulation responsive to the privacy norms of affected populations.

\subsection{Systems Design Applications}
The application of CI surveys to guide systems and product design is covered in detail in our previous work~\cite{noah}. To summarize, device manufacturers can conduct CI surveys to determine whether information collection practices of devices or new features under development will violate consumer privacy norms. This allows modifications during the design process to prevent consumer backlash and public relations debacles.

Applying CI surveys to evaluate privacy regulation can also yield valuable insights for systems research and development. For example, learning that parents value the ability to revoke consent or delete information (Figure~\ref{fig:flows-results}) motivates research into verifiable deletion of cloud data from IoT platforms.
Such insights are especially relevant as neither privacy norms nor regulations are necessarily tied to technical systems feasibility. Discovering that a particular CI parameter value is crucial to privacy norm adherence could launch several research projects developing efficient implementations or correctness proofs. 
We expect future applications of the CI survey method will generate many such results. 

\section{Conclusion}
Increased interest in data privacy has spurred new and updated regulation  around the world. However, there are no widely accepted methods to determine whether this regulation actually aligns with the privacy preferences of those it seeks to protect. 
Here, we demonstrate that a previously developed survey technique~\cite{noah} based on the formal theory of contextual integrity (CI) can be adapted to effectively measure whether data privacy regulation matches the norms of affected populations. 
We apply this methodology to test whether the Children's Online Privacy Protection Act's restrictions on data collection by Internet-connected ``smart'' toys align with parents' norms. We survey 195 parents of children younger than 13 about the acceptability of 1056 smart toy information flows that describe concrete data collection scenarios with and without COPPA restrictions. 

We find that information flows conditionally allowed by COPPA are generally viewed as acceptable by the surveyed parents, in contrast to identical flows without COPPA-mandated restrictions. 
These are the first data indicating the general alignment of COPPA to parents' privacy norms for smart toys. 
However, variations in information flow acceptability across smart toys, information types, and respondent demographics emphasize the importance of detailed contextual factors to privacy norms and motivate further study. 

COPPA is just one of many U.S. and international data privacy regulations. We hope that this work will serve as a template for others to adopt and repeat the CI survey method to study other legislation, allowing for a cross-sectional and longitudinal picture of the ongoing relationship between regulation and social privacy norms. 

\section*{Acknowledgments}
We thank Yan Shvartzshnaider and our survey respondents. This work was supported by the Accenture Fund of the School of Engineering and Applied Science at Princeton University. 

{\normalsize \bibliographystyle{acm}
\bibliography{CI_COPPA}}

\onecolumn
\appendix

\section*{Appendix A: Survey Overview}
\label{app:survey-overview}
\vspace{24pt}

\begin{center}
\fbox{\includegraphics[width=0.7\textwidth]{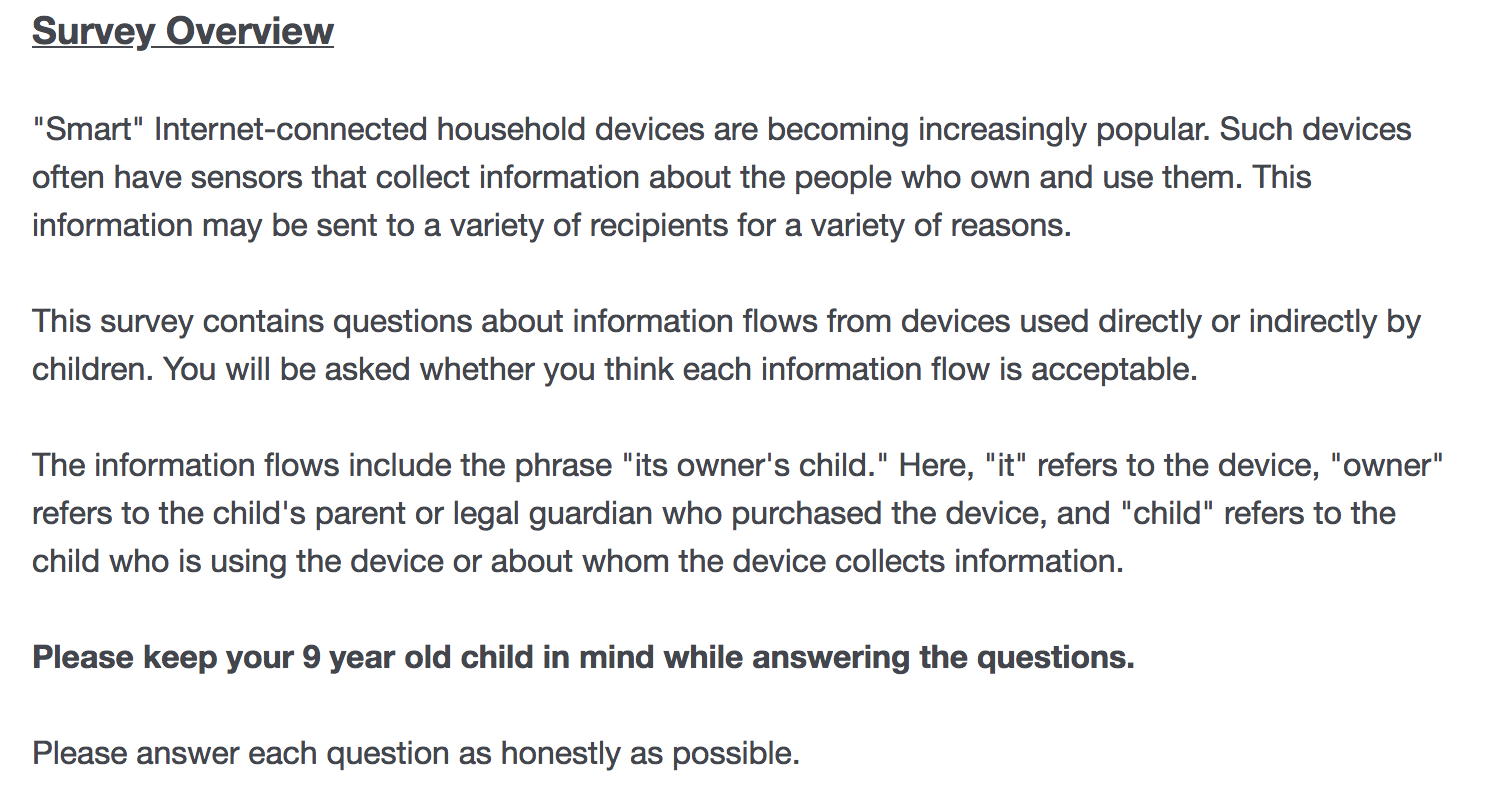}}
\end{center}

\noindent Survey overview shown to participants before contextual integrity information flow questions. Participants are asked to keep one child in mind when answering the survey questions. The age of this child (9 in above example) is selected randomly for each participant from the self-reported ages of each of their children younger than 13.

\clearpage
\section*{Appendix B: Self-Reported Demographics and Technical Background of Survey Respondents}
\label{app:demographics}
\vspace{24pt}

\begin{center}
\begin{tabular}{@{}rlrl@{}}
\toprule
\textbf{Metric}                                          & \multicolumn{1}{l|}{\textbf{Sample}} & \textbf{Metric}                                  & \textbf{Sample} \\ \midrule
                                                         & \multicolumn{1}{l|}{}                &                                                  &                 \\
Female                                                   & \multicolumn{1}{l|}{61\%}            & 18-24 years old                                  & 3\%             \\
Male                                                     & \multicolumn{1}{l|}{39\%}            & 25-34 years old                                  & 31\%            \\
Other/Prefer not to disclose                            & \multicolumn{1}{l|}{-}               & 35-44 years old                                  & 48\%            \\
                                                         & \multicolumn{1}{l|}{}                & 45-54 years old                                  & 13\%            \\
9th, 10th, 11th, 12th - no diploma                       & \multicolumn{1}{l|}{1\%}             & 55-64 years old                                  & 4\%             \\
High school graduate  & \multicolumn{1}{l|}{14\%}            & 65 years or older                                & \textless{}1\%  \\
Some college but no degree                               & \multicolumn{1}{l|}{22\%}            &                                                  &                 \\
Associate degree in college - Vocational                 & \multicolumn{1}{l|}{6\%}             & Has 1 child                                      & 33\%            \\
Associate degree in college - Academic                   & \multicolumn{1}{l|}{4\%}             & Has 2 children                                   & 45\%            \\
Bachelor's degree              & \multicolumn{1}{l|}{30\%}            & Has 3 children                                   & 15\%            \\
Master's degree                                          & \multicolumn{1}{l|}{16\%}            & Has 4 or more children                           & 7\%             \\
Professional school degree                               & \multicolumn{1}{l|}{2\%}             &                                                  &                 \\
Doctorate degree                                         & \multicolumn{1}{l|}{5\%}             & answers based on 0-3 yr old child                & 14\%            \\
                                                         & \multicolumn{1}{l|}{}                & answers based on 4-7 yr old child                & 36\%            \\
Less than \$25,000                                       & \multicolumn{1}{l|}{13\%}            & answers based on 8-12 yr old child               & 50\%            \\
Between \$25,000 and \$50,000                              & \multicolumn{1}{l|}{22\%}            &                                                  &                 \\
Between \$50,000 and \$75,000                              & \multicolumn{1}{l|}{21\%}            & 0-3 hours of internet use per day                & 15\%            \\
Between \$75,000 and \$100,000                             & \multicolumn{1}{l|}{21\%}            & 4-7 hours of internet use per day                & 45\%            \\
Between \$100,000 and \$200,000                            & \multicolumn{1}{l|}{17\%}            & 8-12 hours of internet use per day               & 25\%            \\
More than \$200,000                                      & \multicolumn{1}{l|}{4\%}             & \textgreater{}12 hours of internet use per day   & 14\%            \\
Prefer not to disclose                                   & \multicolumn{1}{l|}{2\%}             &                                                  &                 \\
                                                         & \multicolumn{1}{l|}{}                & Uses a personal computer  & 97\%            \\
Asian                                                    & \multicolumn{1}{l|}{7\%}             & Uses a smartphone                                & 94\%            \\
Black or African American                                & \multicolumn{1}{l|}{11\%}            & Uses a tablet device                             & 78\%            \\
Native Hawaiian or Other Pacific Islander                & \multicolumn{1}{l|}{1\%}             &                                                  &                 \\
White                                                    & \multicolumn{1}{l|}{76\%}            & Owns a smart device*                             & 49\%            \\
White, American Indian or Alaska Native                   & \multicolumn{1}{l|}{\textless{}1\%}  & Does not own a smart device                     & 50\%            \\
White, Asian                                              & \multicolumn{1}{l|}{\textless{}1\%}  & Unsure                                           & \textless{}1\%  \\
White, Black or African American                          & \multicolumn{1}{l|}{1\%}             &                                                  &                 \\
Other                                                    & \multicolumn{1}{l|}{3\%}             & Owns a children's smart device**                 & 33\%            \\
                                                         & \multicolumn{1}{l|}{}                & Does not own a children's smart device         & 66\%            \\
Hispanic                                                 & \multicolumn{1}{l|}{14\%}            & Unsure                                           & 1\%             \\
Not Hispanic                                             & \multicolumn{1}{l|}{85\%}            &                                                  &                 \\
Prefer not to disclose                                   & \multicolumn{1}{l|}{\textless{}1\%}  & Familiar with COPPA                              & 63\%            \\
                                                         & \multicolumn{1}{l|}{}                & Not familiar with COPPA                          & 33\%            \\
                                                         &                                      & Maybe familiar with COPPA                       & 4\%             \\ \bottomrule
\multicolumn{4}{p{6in}}{\footnotesize * Question text: ``Do you own any `smart' (Internet-connected) devices or appliances besides a smartphone, tablet, laptop, or desktop computer?''} \\
\multicolumn{4}{p{6in}}{\footnotesize ** Question text: ``Do you own any `smart' (Internet-connected) devices or appliances used directly or indirectly by children besides a smartphone, tablet, laptop, or desktop computer?''}
\end{tabular}%
\end{center}

\end{document}